\documentclass[twocolumn,letterpaper,aps,prb,showpacs,floatfix,10pt]{revtex4-1}
\usepackage{graphicx}    
\usepackage{hyperref}    
\usepackage{bm}          
\usepackage[sumlimits,intlimits]{amsmath}
\usepackage{amsfonts,amssymb}
\pdfoutput=1
\begin{document}
	
\title{Hopping transport in systems of finite thickness or length}
\author{A. S. Rodin}
\author{M. M. Fogler}
\affiliation{University of California San Diego, 9500 Gilman Drive, La
Jolla, California 92093}

\date{\today}

\begin{abstract}

Variable-range hopping transport along short one-dimensional wires and across the shortest dimension of thin three-dimensional films and narrow two-dimensional ribbons is studied theoretically. Geometric and transport characteristics of the hopping resistor network are shown to depend on temperature $T$ and the dimensionality of the system. In two and three dimensions the usual Mott law applies at high $T$ where the correlation length of the network is smaller than the sample thickness. As $T$ decreases, the network breaks into sparse filamentary paths, while the Mott law changes to a different $T$-dependence, which is derived using the percolation theory methods. In one dimension, deviations from the Mott law are known to exist at all temperatures because of rare fluctuations. The evolution of such fluctuations from highly-resistive ``breaks'' at high $T$ to highly-conducting ``shorts'' at low $T$ is elucidated.


\end{abstract}

\pacs{
72.20.Ee,  
73.63.Nm 
}
\maketitle

\section{Introduction}
\label{sec:Introduction}

Variable-range hopping (VRH) is the mechanism of low-temperature transport common in systems with electron states localized by disorder. Typically, the VRH conductivity $\sigma(T)$ obeys the Mott law,
\begin{equation}
\ln \sigma(T) = -\left(\frac{c_d T_0}{T}\right)^{{1} / (d + 1)},
\quad T_0 = \frac{1}{g a^d}\,,
\label{eqn:G_Mott}
\end{equation}
where $d$ is the space dimension, $a$ is the localization length, $g$ is the density of states, and $c_d$ is a numerical coefficient. Temperature dependence stronger than the Mott law can arise due to electron interactions that deplete $g$ near the Fermi level.~\cite{Shklovskii1984epo} On the other hand, $T$-dependences weaker than the Mott law have also been observed. For example, a power-law scaling,
\begin{equation}
\sigma(T) \sim T^\alpha,
\label{eqn:G_Power_Law}
\end{equation}
has been measured~\cite{Xu1995dih, Yoshida1997tah} in transport across disordered thin films. Such a behavior is thought to originate from hopping along special highly conductive chains of sites.~\cite{Pollak1973not, Tartakovskii1987hco, Glazman1988ita, Levin1988thc} The exponent $\alpha$ scales with the number of hops in the chain. We call this transport mechanism rare-chain hopping (RCH). The RCH has been often discussed in the context of magnetic tunneling junctions, see a recent example in Ref.~\onlinecite{Teixeira2011rtt}.

In the preceding paper~\cite{Rodin2010apl} we have shown that the RCH can also determine the conduction measured in an ensemble of short one-dimensional (1D) wires connected in parallel. In our theory the exponent $\alpha$ depends on the ratio $L / a$, where $L$ is the length of the wires. However, Eq.~\eqref{eqn:G_Power_Law} is only an approximation that holds in a limited range of $T$. Such an apparent power-law behavior has been observed in a number of 1D and quasi-1D systems.~\cite{Zaitsev2000lll, Slot2004odc, Venkataraman2006eti, Zhou2007ode} We demonstrated that the RCH provides a more plausible explanation of these observations than models based on the concept of 1D Luttinger liquid.~\cite{Rodin2010apl}

The present paper is devoted to the crossover from the low-$T$ RCH to
the higher-$T$ Mott law. To our knowledge this problem has not been studied theoretically although similar problems have been examined in the context of VRH transport along the longer dimension of three-dimensional (3D) films~\cite{Shklovskii1975hci} and two-dimensional (2D) strips.~\cite{Raikh1990sei}

We show that depending on space dimension $d$, the Mott law and the RCH represent either two separate, competing contributions to the transport or they succeed one another via a continuous evolution. In the first case, realized in $d = 3$, the logarithmic derivative $d \ln \sigma / d T$ has a sharp change at the crossover point. Systems of dimension $d < 3$ produce the other type of behavior, where a gradual variation of $\ln \sigma(T)$ takes place. Our analysis is most complete in 1D (by which we again mean an array of 1D systems connected in parallel), where we can utilize both numerical and analytical methods. More complicated cases of higher dimensions are studied using qualitative physical considerations.

The paper is organized as follows. Sections~\ref{sec:Model} and \ref{sec:Two} introduce the model and key relations we use to analyze the problem. Sections~\ref{sec:Higher_D} and \ref{sec:1D_systems} deal with $d > 1$ and $d = 1$ systems, respectively. Concluding remarks are given in Sec.~\ref{sec:Conclusions}. Details of the derivations are gathered in the Appendix.

\section{Conductivity from crossing probability}
\label{sec:Model}

We consider localized states (sites) distributed randomly with the uniform average density $g$ in energy-position space. The $x$-coordinates of the states belong to the interval $0 < x < L$. If $d > 1$, the system is assumed to be infinite along the remaining $d - 1$ coordinates. The $x = 0$ side of the system is the source electrode and the $x = L$ side is the drain electrode. We wish to compute the conductivity $\sigma(T, L)$ between the source and the drain.

We adopt the Miller-Abrahams resistor network model~\cite{Miller1960ica}
in which the resistance between any two sites is given by
\begin{equation}
R_{i j} = R_0 e^{u_{i j}}\,, \quad u_{ij} = \frac{2 r_{i j}}{a} + \frac{|\varepsilon_i| + |\varepsilon_j| + |\varepsilon_i - \varepsilon_j|}
     {2 T}\,,
\label{eqn:u}
\end{equation}
where $r_{i j}$ is the distance between the sites and $\varepsilon_i$ is the energy of state $i$ measured from the Fermi level. The coefficient $R_0$ should have some power-law $T$-dependence determined by the electron-phonon coupling. However, we will ignore it and treat $R_0$ as a constant.

The next approximation is to view our system as an array of hypercubes of length $L$ and cross-section area $L^{d - 1}$ connected in parallel. (For an array of $d = 1$ wires this approximation is exact.) It yields the relation between the conductivity and the ensemble-averaged conductance $G$ of a single hypercube:
\begin{equation}
\sigma(T, L) = L^{2 - d} \langle G \rangle\,.
\label{eqn:sigma}
\end{equation}
In order to estimate $\langle G \rangle$ we use the percolation theory argument. We assume that the conductivity of a given hypercube is dominated by an optimal subnetwork, which is constructed as follows.~\cite{Ambegaokar1971hci, Shklovskii1984epo} Each pair of sites with $u_{ij} \leq u$ is considered a connected bond; otherwise, it is a broken bond. Gradually increasing $u$, one reaches some value $u_c$ where for the first time the source and the drain become joined by a path of connected bonds. The dominant subnetwork is obtained increasing $u$ up to $u_c + 1$ or so. Further increase in $u$ is assumed not to lead to a significant growth of the conductivity because the added resistors would be shunted by those already present in the circuit.

%
%
\begin{figure}
  \includegraphics[width=2.6in]{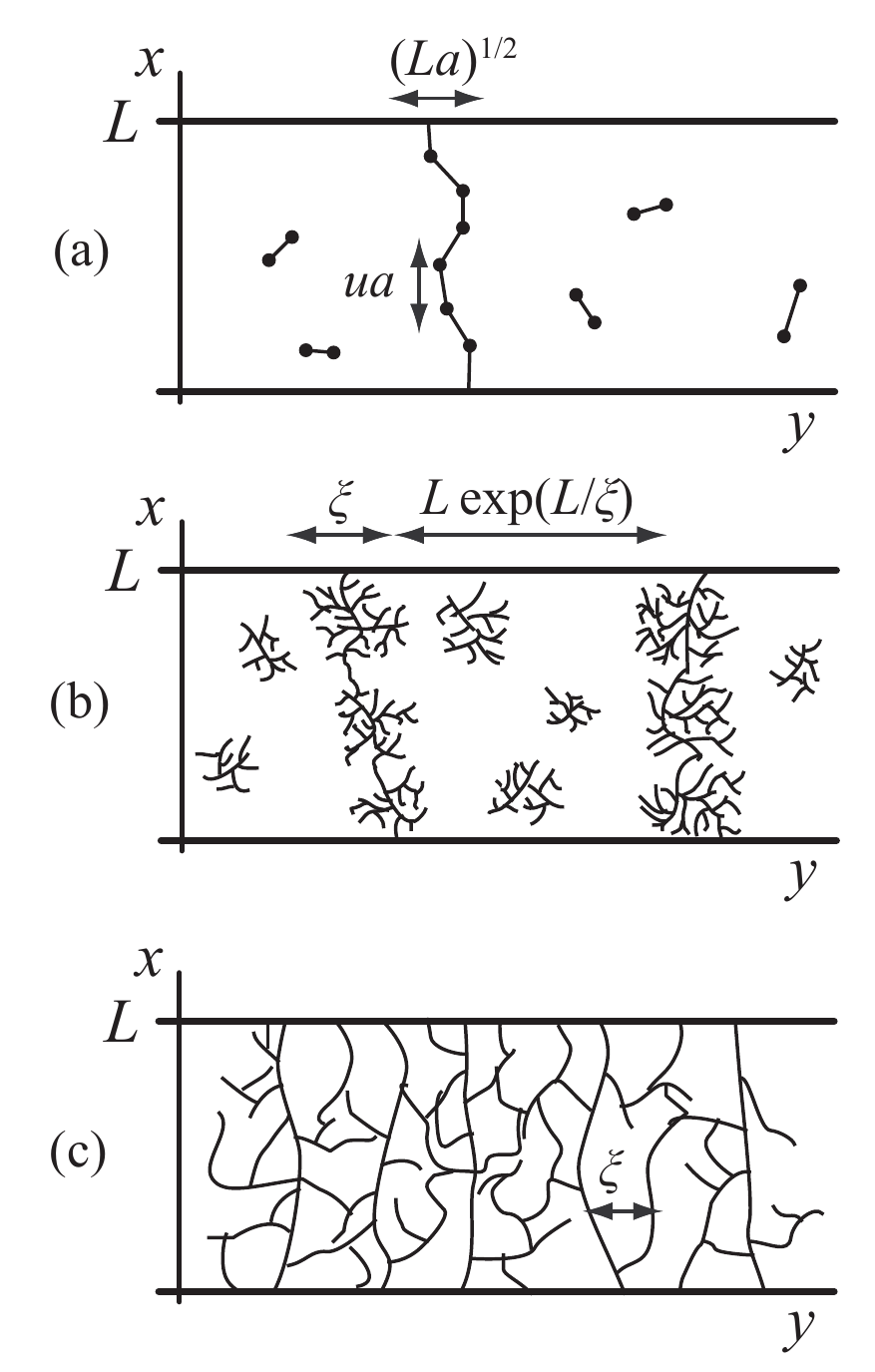}
  \caption{\label{fig:Network} Evolution of a 2D network with increasing $T$ (see Sec.~\ref{sec:Higher_D}). The network progresses from independent conducting strands to an interconnected grid, see main text.}
\end{figure}

For each disorder realization, the critical $u$ is a random number. Its statistical properties are encoded in the so-called crossing probability~\cite{Berlyand1995tpb, Hovi1996sau, Newman2001fmc} $P(u)$, which is a cumulative distribution function of $u$. Thus, $P(u)$ gives the probability of having a connected path (the spanning cluster) between the source and the drain via the bonds $u_{i j} \leq u$. (Accordingly, the derivative of $P(u)$ is the distribution function of the percolation thresholds in a finite-size hypercube, the earliest study of which was carried out in Ref.~\onlinecite{Levinshtein1976trb}.) Understanding the geometry of the spanning cluster is a key to calculating both $P(u)$ and the average conductance $\langle G \rangle$.

In general, the cluster consists of the current-carrying backbone and the dead-ends. We can imagine two limits. The backbone can be made of a single filament, as illustrated in Fig.~\ref{fig:Network}(a) and (b), or it can look like a $d$-dimensional network with a correlation length (characteristic size of the cells) $\xi \ll L$, see Fig.~\ref{fig:Network}(c). In the latter case, which is possible only if $d > 1$, we find
\begin{equation}
\begin{split}
\langle G \rangle &\sim (L / \xi)^{d - 2} \int d u P^\prime(u) R_0^{-1} e^{-u}\\
&= (L / \xi)^{d - 2} R_0^{-1} \int d u P(u) e^{-u}\,,
\end{split}
\label{eqn:G_avr_I}
\end{equation}
The second line in Eq.~\eqref{eqn:G_avr_I} is obtained integrating by parts. Substituting it into Eq.~\eqref{eqn:sigma}, we obtain, with the same accuracy,
\begin{equation}
\sigma \sim R_0^{-1} \max\limits_{u} \left\{
                     \xi^{2 - d}(u) P(u) e^{-u}
                     \right\}\,.
\label{eqn:G_avr_II}
\end{equation}
Similarly, in the case of a single filament of a characteristic transverse dimension $L_\perp$, we can estimate the conductivity as follows:
\begin{equation}
\sigma \sim R_0^{-1} \max\limits_{u} \left\{
                     L L_\perp^{1 - d}(u) P(u) e^{-u}
                     \right\}\,.
\label{eqn:G_avr_III}
\end{equation}
The brunt of the remaining work is to compute $P(u)$ and $L_\perp$ (or $\xi$) as a function of $T$, $L$, and $d$.

\section{Two main regimes}
\label{sec:Two}

As discussed in Sec.~\ref{sec:Introduction}, there are two principal regimes of VRH, the Mott law and the RCH. In this section we show how they follow from our Eqs.~\eqref{eqn:G_avr_I}-\eqref{eqn:G_avr_III}.

\subsection{The Mott law}
\label{sec:Critical}

To see how Eq.~\eqref{eqn:G_avr_I} leads to the Mott law let us recall that according to the percolation theory, in the limit $L \to \infty$, the crossing probability $P(u)$ approaches a step-function, $P(u) \to \Theta(u - u_c)$; therefore,
\begin{equation}
 \ln \sigma \simeq -u_c\,. 
\label{eqn:sigma_from_u_c}
\end{equation}
The integral in Eq.~\eqref{eqn:G_avr_I} is dominated by the interval
\begin{equation}
u_c<u\lesssim u_c+1\,.
\label{eqn:u_interval}
\end{equation}
To calculate the threshold value $u_c$ we proceed as follows. From Eq.~\eqref{eqn:u} we see that for a given $u$ we need to consider only sites with energies $-T u < \varepsilon < T u$. These sites have the average coordination number
\begin{equation}
C(u) = \frac{\int_{-T u}^{T u} g d\varepsilon_j \int_{-T u}^{T u} g d \varepsilon_i \int d^d r_{i j} \Theta(u - u_{i j})}
{\int_{-T u}^{T u} g d\varepsilon}\,,
\label{eqn:B1}
\end{equation}
which scales with $u$ as
\begin{align}
C(u) &= \beta_d \left({u} / {u_M}\right)^{d + 1}\,
\label{eqn:B2}\\
u_M &\equiv (2 T_0 / T)^{1 / (d + 1)}\,,
\label{eqn:u_M}
\end{align}
where $\beta_1 = 1/2$, $\beta_2 = \pi/8$, and $\beta_3 = \pi/20$. The percolation threshold is known to be $C_c \approx 3$ for $d = 2, 3$. Hence, the threshold value is $u_c = (C_c / \beta_d)^{1 / (d + 1)} u_M$. Substituting this into Eq.~\eqref{eqn:sigma_from_u_c}, we recover the Mott law [Eq.~\eqref{eqn:G_Mott}] and determine the numerical factor therein to be $c_d = 2 C_c / \beta_d$.

The geometry of the critical subnetwork in the Mott regime is of course more complicated~\cite{Shklovskii1984epo} than what is sketched in Fig.~\ref{fig:Network}(c). This subnetwork is a self-similar fractal object on the spatial scales $r_M \ll r \ll \xi$. Here
\begin{equation}
 r_M = a u_M / 2
\label{eqn:r_M}
\end{equation}
is the typical hopping length,
\begin{equation}
\xi = r_M u_M^{\nu_d} \sim a u_M^{\nu_d + 1}\,,
\label{eqn:xi}
\end{equation}
is the correlation length, and
\begin{subequations}
\label{eqn:nu}
\begin{align}
\nu_2 &= 4 / 3       \quad (d = 2)\,,
\label{eqn:nu_2}\\
\nu_3 &\approx 7 / 8 \quad (d = 3)
\label{eqn:nu_3}
\end{align}
\end{subequations}
are the percolation theory exponents.~\cite{Shklovskii1984epo, Isichenko1992pst} Equation~\eqref{eqn:xi} for $\xi$ follows from the general formula
\begin{equation}
\xi \sim r_M |\epsilon|^{-\nu_d}\,,
\label{eqn:xi_from_eps}
\end{equation}
where
\begin{equation}
 \epsilon = u / u_c - 1 
\label{eqn:epsilon}
\end{equation}
is the fractional distance to the percolation point. For $u = u_c+1$ [cf.~Eq.~\eqref{eqn:u_interval}], we have $\epsilon \sim 1 / u_M$, leading to Eq.~\eqref{eqn:xi}.

The above derivation applies if $d > 1$. In 1D the percolation threshold does not exist in the $L \to \infty$ limit because the infinite spanning cluster is invariably broken apart by fluctuations. We discuss this special case in Sec.~\ref{sec:1D_systems}.

\subsection{Rare-chain hopping}
\label{sec:RCH}

It is easy to see that for $\xi$ given by Eq.~\eqref{eqn:xi}, the condition $L \gg \xi$ necessary for validity of the Mott law can be satisfied only at high enough temperature. At lower $T$, the network geometry must be different and deviations from the Mott law should appear.~\footnote{The criterion $\xi \sim L$ also determines the onset of finite-size effects in transport along the longer dimension of a 3D film~\cite{Shklovskii1975hci} or a 2D strip.~\cite{Raikh1990sei}} For the lowest $T$ we expect the RCH regime. Let us now rederive the corresponding conductivity~\cite{Pollak1973not, Tartakovskii1987hco, Glazman1988ita, Levin1988thc} from our formalism.

The connectivity of the chain of $N$ sites is determined by the probability of forming $N$ consecutive bonds. Accordingly, $P(u) \sim [C(u)]^N$. The lower bound on $N$ is $2 L /a u$ because the length of each hop is less than or equal to
\begin{equation}
R = u a / 2\,.
\label{eqn:R}
\end{equation}
Since $C \ll 1$, the optimal number $N_*$ of sites must be close to this bound; hence,
\begin{equation}
\ln P(u) \simeq -\frac{2 L}{a u}\, \ln \left[\left(\frac{u_M}{u}\right)^{d + 1}\right]\,,
\quad u \ll u_M\,.
\label{eqn:P_low_u}
\end{equation}
Substituting this into Eq.~\eqref{eqn:G_avr_III}, we find the optimal $u$ to be, with logarithmic accuracy
\begin{gather}
u_* \simeq \left[
           \frac{2 (d + 1) L}{a}\, \ln \left(
           \frac{u_M}{u_{\text{RCH}}}\right)
           \right]^{1/2}\,,
\label{eqn:u_*_RCH}\\
u_{\text{RCH}} \equiv \sqrt{\frac{2 L}{a}}\,.
\label{eqn:u_RCH}
\end{gather}
For the conductivity, we find
\begin{equation}
\ln\sigma \simeq -\left[
           \frac{8 L}{a}\, \ln \left(\frac{T_0}{T}\,  
                               \frac{1}{u_{\text{RCH}}^{d+1}}\right)
           \right]^{1/2}\,,
\label{eqn:sigma_RCH}
\end{equation}
in agreement with the previous work.~\cite{Tartakovskii1987hco} Equation~\eqref{eqn:sigma_RCH} represents a very slow $T$-dependence compared to the Mott law. If the accessible range of $T$ is limited, as is often the case in experiments, such a dependence can be easily confused with a power-law, Eq.~\eqref{eqn:G_Power_Law}. In fact, Eq.~\eqref{eqn:sigma_RCH} can be written as
\begin{equation}
\sigma \sim T^{N_*}\,.
\end{equation}
Therefore, the exponent $\alpha$ in Eq.~\eqref{eqn:G_Power_Law} is essentially the optimal number of sites $N_* \propto \ln^{-1/2}[(T / T_0) u_{\text{RCH}}^{d + 1}]$ in the chain.~\cite{Glazman1988ita, Xu1995dih, Rodin2010apl}

The remainder of the paper is devoted to analyzing the crossover between the Mott law and the RCH.

\section{2D and 3D systems}
\label{sec:Higher_D}

We make a key observation that at finite $L$, function $P(u)$ is nonvanishing even at $u < u_c$ due to some disorder realizations that percolate ``early.'' This creates an exponential tail of $P(u)$, which competes with the factor $e^{-u}$ in Eq.~\eqref{eqn:G_avr_III}. As a result, the optimal $u$ can be pushed below the threshold, $u_* < u_c$, implying that the transport is governed by subcritical percolation, $\epsilon < 0$.

The behavior of the crossing probability near the percolation threshold
$0 < -\epsilon \ll 1$ can be understood qualitatively as follows. Correlation length $\xi$ in Eq.~\eqref{eqn:xi_from_eps} represents the characteristic size of the largest connected clusters (i.e., the largest clusters among those that are not yet exponentially rare~\cite{Isichenko1992pst}). This is different from the geometrical meaning of $\xi$ in the supercritical regime $u > u_c$ where it is the characteristic size of the voids in the network, see Eq.~\eqref{eqn:u_interval} and Fig.~\ref{fig:Network}(c). In the subcritical regime, the spanning cluster appears when $L / \xi$ independent clusters of size $\xi$ each join together by chance, forming a conducting pathway. Therefore, the crossing probability can be estimated as~\cite{Hovi1996sau, Newman2001fmc} $P(u) \sim e^{-L / \xi}$. Combined with Eqs.~\eqref{eqn:xi_from_eps} and \eqref{eqn:epsilon}, this estimate yields~\footnote{in 2D case, this coincides with Eq.~(9) of Ref.~\onlinecite{Raikh1990sei}.}
\begin{equation}
\ln P(u) \simeq -\frac{L}{\xi} \sim -\frac{L}{r_M}\, \left(\frac{u_c - u}{u_c}\right)^{\nu_d}\,.
\label{eqn:P_high_u}
\end{equation}
This implies
\begin{equation}
P^\prime(u) \propto (u_c - u)^{\nu_d - 1}\,.
\label{eqn:P_Derivative}
\end{equation}
Hence, the behavior of $P(u)$ near $u_c$ depends on whether $\nu_d$ is larger or smaller than unity. In 2D, we have $\nu_2 > 1$, and so the derivative vanishes. In 3D, the opposite inequality $\nu_3 < 1$ holds, and so $P^\prime(u_c)$ diverges. This dichotomy is the reason for the different manner in which the Mott law transitions to the RCH regime in the two cases.

Consider the 2D case first. Per Eq.~\eqref{eqn:G_avr_II} the conductivity is determined by the maximum of $e^{-u} P(u)$. Using Eq.~\eqref{eqn:P_high_u}, we find that it is reached at $u_*$ such that
\begin{equation}
u_c - u_* = ({T_P} / {T})^\beta\,,
\quad T_{\text{RCH}} < T < T_P\,,
\label{eqn:u_max_high_T}
\end{equation}
where
\begin{equation}
\beta = \frac{\nu_d + 1}{(d + 1)(\nu_d - 1)} = \frac{7}{3}\,,
\quad d = 2\,,
\label{eqn:beta}
\end{equation}
and the characteristic temperatures $T_{\text{RCH}}$ and $T_P$ are
\begin{align}
T_P &\sim T_0 \left(\frac{a}{L}\right)^{(d + 1) / (\nu_d + 1)} =
T_0 \left(\frac{a}{L}\right)^{{9}/{7}}\,,
\label{eqn:T_P}\\
T_{\text{RCH}} &\sim T_0 \left(\frac{a}{L}\right)^{(d + 1) / 2} =
T_0 \left(\frac{a}{L}\right)^{{3}/{2}}\,.
\label{eqn:T_RCH}
\end{align}
$T_P$ is temperature below which the $T$-dependence of the conductivity and the network geometry start to deviate from what is found in an infinite sample, cf.~Sec.~\ref{sec:Critical} and Fig.~\ref{fig:Network}(c). The Mott law acuires the correction term as follows:
\begin{equation}
\ln \sigma = -u_c + \frac{\nu_d - 1}{\nu_d}\, (u_c - u_*)\,.
\label{eqn:sigma_2D}
\end{equation}
Formula~\eqref{eqn:xi} for $\xi$ is replaced by $\xi \sim L / (u_c - u_*) \ll L$ and the VRH network now consists of well separated filaments, see  Fig.~\ref{fig:Network}(b).

As $T$ decreases further and reaches $T_{\text{RCH}}$, the optimal $u$ becomes equal to $u_{\text{RCH}}$ [Eq.~\eqref{eqn:u_RCH}]. At this point the distance from the critical point $u_c - u_*$ becomes comparable to $u_c$ itself. The cluster size $\xi$ shrinks down to the elementary hopping length $R$, which signifies the crossover to RCH, see Sec.~\ref{sec:RCH}.

\begin{figure}
  \includegraphics[width=3.0in]{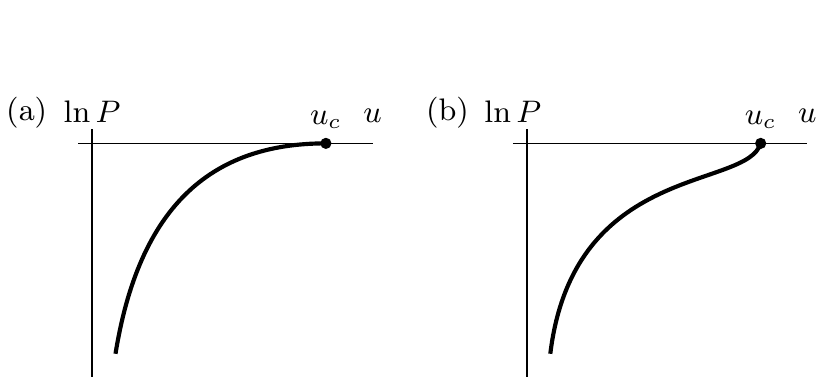}
\caption{Crossing probability $P(u)$. The qualitative difference of the (a) 2D and (b) 3D cases in the vicinity of $u_c$ is apparent.}
\label{fig:PDF_HighD}
\end{figure}

\begin{figure}%
  \includegraphics[width=2.6in]{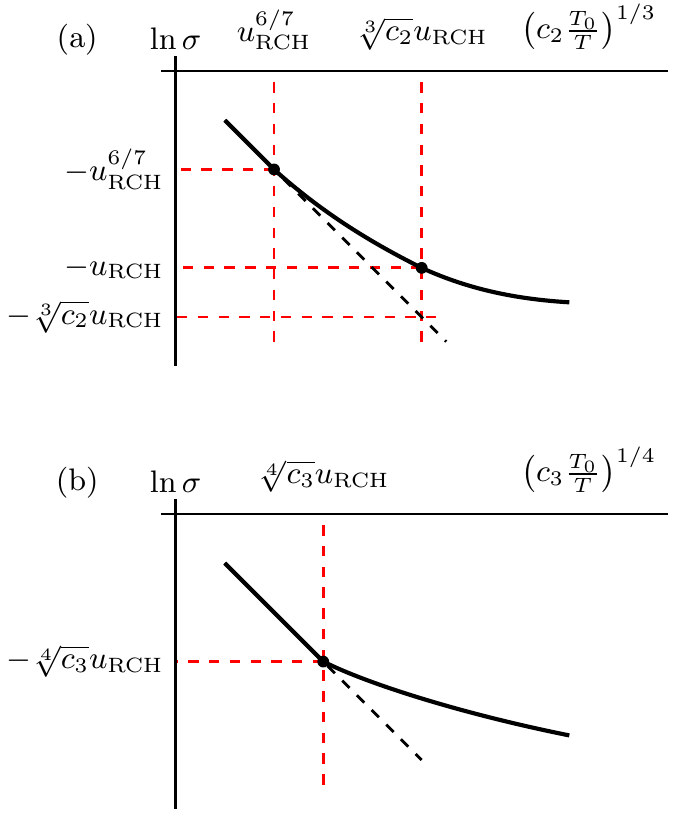}
\caption{Conductivity as a function of temperature.  The tilted dashed line shows the extrapolation of the Mott law past its region of validity for contrast. (a) In 2D, there is an intermediate regime smoothly interpolating between the Mott law and the RCH. (b) In 3D, the transition between the Mott and RCH regimes is abrupt.}%
\label{fig:CondHighD}%
\end{figure}

In 3D, the optimal $u_*$ has a different $T$-dependence. Due to positive concavity (the second derivative) of the $\ln P(u)$ curve near $u_c$, see Fig.~\ref{fig:PDF_HighD}, the maximum of $e^{-u} P(u)$ stays at $u_c$ at all $T > T_{\text{RCH}}$. Temperature scale $T_P$ plays no role and Eq.~\eqref{eqn:u_max_high_T} does not apply. The Mott law continues to be valid. However, as soon as $T$ drops below $T_{\text{RCH}}$, parameter $u_*$ gets suddenly reduced by some numerical factor and at still lower $T$ we have Eq.~\eqref{eqn:u_*_RCH}. In other words, in 3D the transition from the Mott law to RCH is abrupt, see Fig.~\ref{fig:CondHighD}(b).

\section{Quasi-1D systems}
\label{sec:1D_systems}

\subsection{Toy model}

Since the percolation threshold does not exist in 1D, function $P(u)$ for high $u$ has a somewhat different form compared to $d > 1$. The peculiarities of the 1D case are well illustrated by the toy model in which all the site energies are equal to zero. The connected sites are those that are separated by distance less than $R$ [Eq.~\eqref{eqn:R}].

The crossing probability can be computed exactly:
\begin{equation}
P = 1 + \sum\limits_{1 \leq k \leq l} \frac{(-1)^k}{k!}\, e^{-\rho k}
[\rho(l - k)]^{k - 1} (k - k \rho + l \rho)\,,
\label{eqn:P_sol}
\end{equation}
where $l =  L / R$, $\rho = g R$, and the density of states $g$ is redefined to be simply the density in coordinate space. For large $l$ this expression can be approximated by
\begin{equation}
\ln P \sim (s_0 - \rho) l\,,
\label{eqn:Approx_Misha}
\end{equation}
where $s_0$ is the real root of the equation
\begin{equation}
k(s) \equiv \frac{1 - e^{-s}}{s} = \frac{1}{\rho}\,,
\label{eqn:k}
\end{equation}
cf.~Appendix~\ref{sec:Interp_Form}. For large $\rho$, this yields
\begin{equation}
\ln P \simeq -l \rho e^{-\rho}\,,
\quad \rho \gg 1\,.
\label{eqn:High_rho_toy_model}
\end{equation}
For small $\rho$, we have $s_0 \sim \ln \rho$, which gives
\begin{equation}
\ln P \simeq -l \ln \frac{1}{\rho}\,,
\quad \rho \ll 1\,.
\label{eqn:Low_rho_toy_model}
\end{equation}
These results have a simple interpretation. For high $\rho$, the sites are very dense. However, empty segments of length larger or equal to $R$ may also appear by chance. Deviations of $P$ from unity are due to these rare disruptions --- ``breaks.''  The probability of having no sites in a segment of length $R$ is
\begin{equation}
              p_{\text{toy}} = e^{-\rho}. 
\label{eqn:p_toy}
\end{equation}
The average distance $1 / (p_{\text{toy}} g)$ between such breaks defines the average length $\xi$ of connected clusters. Using $\ln P = -L / \xi$, as in Eq.~\eqref{eqn:P_high_u}, we recover Eq.~\eqref{eqn:High_rho_toy_model}. On the other hand, for small $\rho$, the sites are very dilute, so we have an analog of the RCH regime. We expect $P \sim C^N$, similar to Sec.~\ref{sec:RCH}. Approximating the number of hops $N$ by $l$ and the average coordination number $C$ by $\rho$, we arrive at Eq.~\eqref{eqn:Low_rho_toy_model}. (We assume that $N$ is large and ignore its fractional part in this heuristic derivation.)
Note that Eqs.~\eqref{eqn:High_rho_toy_model} and \eqref{eqn:Low_rho_toy_model} match at $\ln \rho \approx -1$, where the cluster size shrinks to the elementary size $R$. In this respect, our toy model is similar to the 2D case (Sec.~\ref{sec:Higher_D}).

Next, we show that the having random energies in addition to random $x$-coordinates does not qualitatively change this physical picture.

\subsection{Analytical results for 1D}

The model of 1D VRH in which both the coordinates $x$ and energies $\varepsilon$ of the sites are random has been studied extensively in prior literature. It has been shown that a number of rigorous analytical results can be obtained in the limits of either high or low $T$ where $\sigma$ is dominated by rare events. Below we rederive these results in a unified manner, which enables us to elucidate the crossover between them.

%
\begin{figure}
  \includegraphics[width=3in]{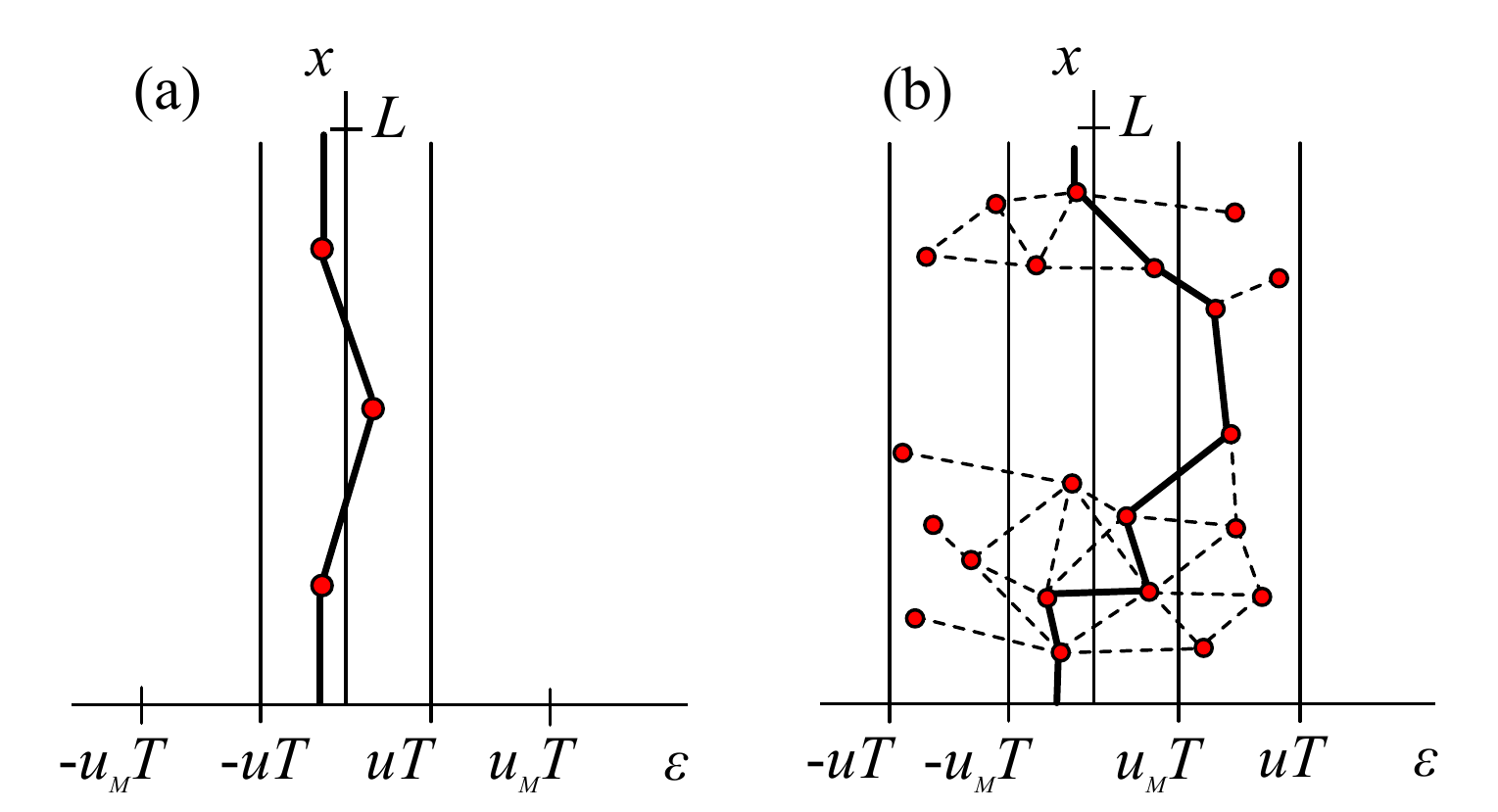}
  \caption{\label{fig:Path_Comparison} VRH network in 1D. (a) At low $T$, the network is made of a few approximately equidistant hops. The resistance of each link $e^u$ is much smaller than the typical one $e^{u_M}$. This structure is common to all $d$, cf.~Fig.~\ref{fig:Network}(a). (b) At high $T$, most of the resistors in the network (dashed lines) have typical values and form clusters. However, unlike the $d > 1$ case [Fig.~\ref{fig:Network}(b)], the clusters are connected together by $u \gg u_M$ ``breaks'' that straddle regions empty of the hopping sites. The solid line shows the path of the least total resistance.}
\end{figure}

\begin{figure}
\includegraphics[width=2.6in]{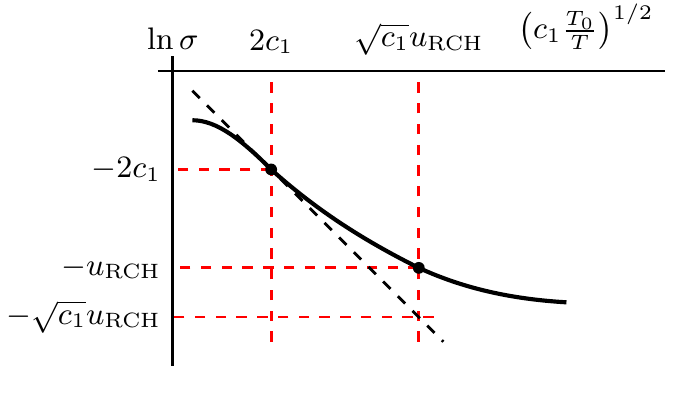}
\caption{\label{fig:1D} Transport regimes in the quasi-1D case. The lowest $T$ region (to the right of the second dot on the curve) is the RCH regime, Eq.~\eqref{eqn:sigma_RCH}. The intermediate $T$-regime (between the two dots)
is described by Eq.~\eqref{eqn:High_G}. It transitions smoothly into activated transport at high $T$. Near the inflection point (the leftmost dot) the Mott law, Eq.~\eqref{eqn:G_Mott}, is realized.}
\end{figure}

Recall that each link of the hopping network is characterized by a dimensionless number $u_{i j}$. As in the toy model, we can talk about ``breaks,'' by which we mean links with $u_{i j}$ much larger than the typical value $u_M$. At high $T$ the transport is dominated by these rare breaks,~\cite{Kurkijarvi1973hci, Lee1984vrh, Raikh1989fot, Raikh1990sei, Ruzin1991fso} see Fig.~\ref{fig:Path_Comparison}(b). The breaks straddle regions of size $R = u a / 2$ and $2 u T$ in $x$ and $\varepsilon$ directions, respectively, empty of hopping sites. The probability of having a break with $u_{i j} > u$ can be computed considering how the area and the perimeter of such regions scale with $u$. The result is~\cite{Raikh1989fot, Ruzin1991fso}
\begin{equation}
p(u) \sim \exp\left(-\frac{u^2}{u_M^2} + 2 B \frac{u}{u_M}\right)\,,
\label{eqn:P_Diamond}
\end{equation}
where~\cite{Rodin2009nso} $B \approx 0.9$. The average size of the connected cluster is therefore $\xi = r_M / p(u)$ and the crossing probability is, similar to Eqs.~\eqref{eqn:P_high_u} and \eqref{eqn:High_rho_toy_model},
\begin{equation}
\ln P(u) \sim -\frac{L}{\xi} \sim -\frac{2 L}{u a} \exp\left(-\frac{u^2}{u_M^2} + 2 B \frac{u}{u_M}\right)\,.
\label{eqn:PL_1D_high_u}
\end{equation}
For such $P(u)$, the maximum of the right-hand side of Eq.~\eqref{eqn:G_avr_III} is reached at
\begin{equation}
u_* \simeq u_M \ln^\frac{1}{2} \left( \frac{u_{\text{RCH}}^2}{u_M^2} \right)\,,
\quad u_M \ll u_{\text{RCH}}\,.
\label{eqn:u_*}
\end{equation}
Accordingly, the conductivity $\sigma$ is given by~\footnote{The argument of the logarithmic factor on the right-hand side is reduced by $\sqrt{T /
T_0}$ compared to Eq.~(7) of Ref.~\onlinecite{Raikh1989fot} for the typical conductance. This is because Eq.~\eqref{eqn:High_G}, equivalent to Eq.~(13) of Ref.~\onlinecite{Rodin2010apl}, represents not the typical but the ensemble-averaged conductivity, which is enhanced by the better conducting members of the ensemble.}
\begin{equation}
\ln \sigma(T) \simeq -\left[\frac{2 T_0}{T}\, \ln\left(\frac{L}{a}\, \frac{T}{T_0}\right)\right]^{1/2}\,.
\label{eqn:High_G}
\end{equation}
It was shown in Ref.~\onlinecite{Raikh1989fot} that this equation has the upper limit of validity, $T = T_0 / (4 c_1)$, where $c_1$ is defined by
\begin{equation}
c_1 = 2\ln u_{\text{RCH}}^2 = 2\ln \left(\frac{2 L}{a}\right)\,.
\label{eqn:c_1}
\end{equation}
At still higher $T$ the conductivity is described by the activation law,~\cite{Kurkijarvi1973hci, Raikh1989fot} $\ln \sigma(T) \simeq -T_0 / (2 T)$.
This implies that Mott law is, strictly speaking, invalid in 1D. However, if the accessible range of temperature is narrow, the deviations from the Mott law may not be readily apparent. These deviations are the smallest precisely
near the point of crossover to the activation law because
the dependence of $\ln \sigma(T)$ on $u_M \propto T^{-1/2}$ must have an inflection point there. In this sense, we can argue that the Mott law is confined to a narrow vicinity of $T = T_0 / (4c_1)$, while Eq.~\eqref{eqn:High_G} describes an intermediate subcritical percolation regime, similar to Eq.~\eqref{eqn:sigma_2D} in 2D. We may expect that this intermediate regime should cross over to the RCH when the cluster size $\xi$ becomes of the order of the average hopping length $r_M$. Indeed, this criterion is satisfied at $u_M \sim u_{\text{RCH}}$, see Fig.~\ref{fig:1D}.

%

For $u < u_{\text{RCH}}$, the behavior of $P(u)$ is determined by the RCH, see Fig.~\ref{fig:Path_Comparison}(a). Adopting the method of Ref.~\onlinecite{Tartakovskii1987hco} to the 1D case, we obtain
the result conceptually similar to Eq.~\eqref{eqn:Low_rho_toy_model}:
\begin{equation}
\ln P(u) \simeq \frac{2 L}{a}\,c_0(u)\,.
\label{eqn:P_approx}
\end{equation}
Function $c_0(u)$ in this equation is defined in Appendix~\ref{sec:Low_u}. For $u \ll u_M$, it has the asymptotic form
\begin{equation}
c_0(u)
\simeq -\frac{1}{u} \ln\left(\frac{u_M^2}{u^2}\right)\,,
\label{eqn:c_0_result}
\end{equation}
so that the crossing probability is given by
\begin{equation}
\ln P(u) \simeq -\frac{2 L}{u a}\,
\ln\left(\frac{u_M^2}{u^2}\right)\,.
\label{eqn:Crossing_low_rho_1D}
\end{equation}
As in our toy model, $P$ is the product of the number of hops $N = 2 L / (u a)$ and a logarithmic factor. Equation~\eqref{eqn:Crossing_low_rho_1D} is also in agreement with the general RCH formula~\eqref{eqn:P_low_u}, and so it leads to Eq.~\eqref{eqn:sigma_RCH} within the leading logarithmic approximation. In principle, Eq.~\eqref{eqn:P_approx} goes to the next order in this approximation. In practice, the logarithms involved are never truly large. This motivates us to additionally examine the problem by numerical simulations, which are described in the last part of this Section.

\subsection{Numerical results for 1D}

To compute $P(u)$ numerically we proceed as follows. First, we generate an ensemble of 1D systems for each $L$ and $T$ of interest. For every realization we find $u_c$ --- the minimum value of $u$ at which the path traversing the system is formed. This $u_c$ is a random number between $0$ and $2 L / a$. To find it we start from some initial guess $\sim L / a$ and then fine-tune it by iterative bisection. The connectivity of the network for a given $u$ is determined by the algorithm similar to one used in our previous work.~\cite{Rodin2009nso, Rodin2010apl} Finally, the histogram of $u_c$ gives us the crossing probability $P(u)$.

%
\begin{figure}[b]
  \includegraphics[width=2.7in]{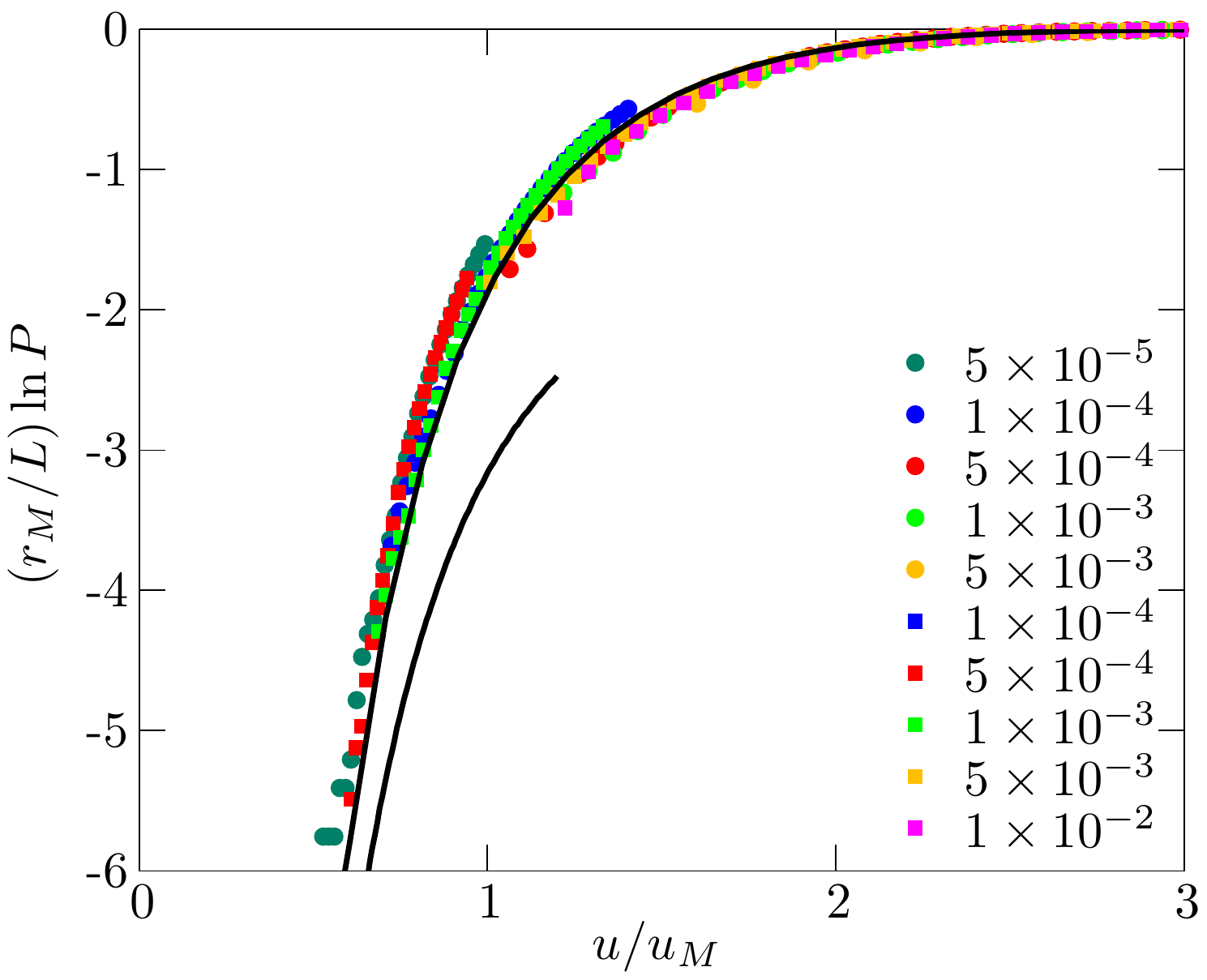}
  \caption{\label{fig:Numerics} (Color online) Numerical results for $P(u)$ in 1D. Different colors represent $T$ ranging from $5\times10^{-5}$ to $5\times10^{-3}$ for $L = 100a$ (circles) and from $10^{-4}$ to $10^{-2}$ for $L = 30a$ (squares), with the temperature unit being such that $T_0 = 3 / 4$. The solid lines are analytic approximations, see main text.}
\end{figure}

In Fig.~\ref{fig:Numerics} we show that when our 
results for $P(u)$ are plotted as $(r_M / L) \ln P(u)$ versus $u / u_M$ they collapse on a common master curve. The deviation from the master curve at high $u$ come from the fact that the number of hops $N \sim 2$ is no longer large. The deviations at low $u$ are due to poor statistics of these low probability events.

We considered two analytical approximations of our numerical results. First, aiming to utilize the analogy to the exactly solvable toy model, we fitted the master curve in terms of function $P$ defined by Eq.~\eqref{eqn:P_sol}. The fit (shown by the longer curve in Fig.~\ref{fig:Numerics}) was generated by setting
\begin{equation}
\rho = \frac{1}{1 + 2B \frac{u_M}{u}}\frac{u^2}{u_M^2}
\end{equation}
designed to produce $p \simeq p_{\text{toy}}$ at $u \gg u_M$, cf.~Eqs.~\eqref{eqn:p_toy} and \eqref{eqn:P_Diamond}.

The second approximation is the parameter-free Eq.~\eqref{eqn:P_approx}, which simultaneously represents the leading asymptotic result for $P(u)$ at low $u$ and the strict lower bound at any $u$, see Appendix~\ref{sec:Low_u}. It is shown by the shorter solid curve in Fig.~\ref{fig:Numerics}. While our numerical results converge to this curve at lowest $u$, we were unable to achieve an accurate match because of prohibitive computational cost needed to acquire statistics in this region.

%
\begin{figure}
  \includegraphics[width=2.6in]{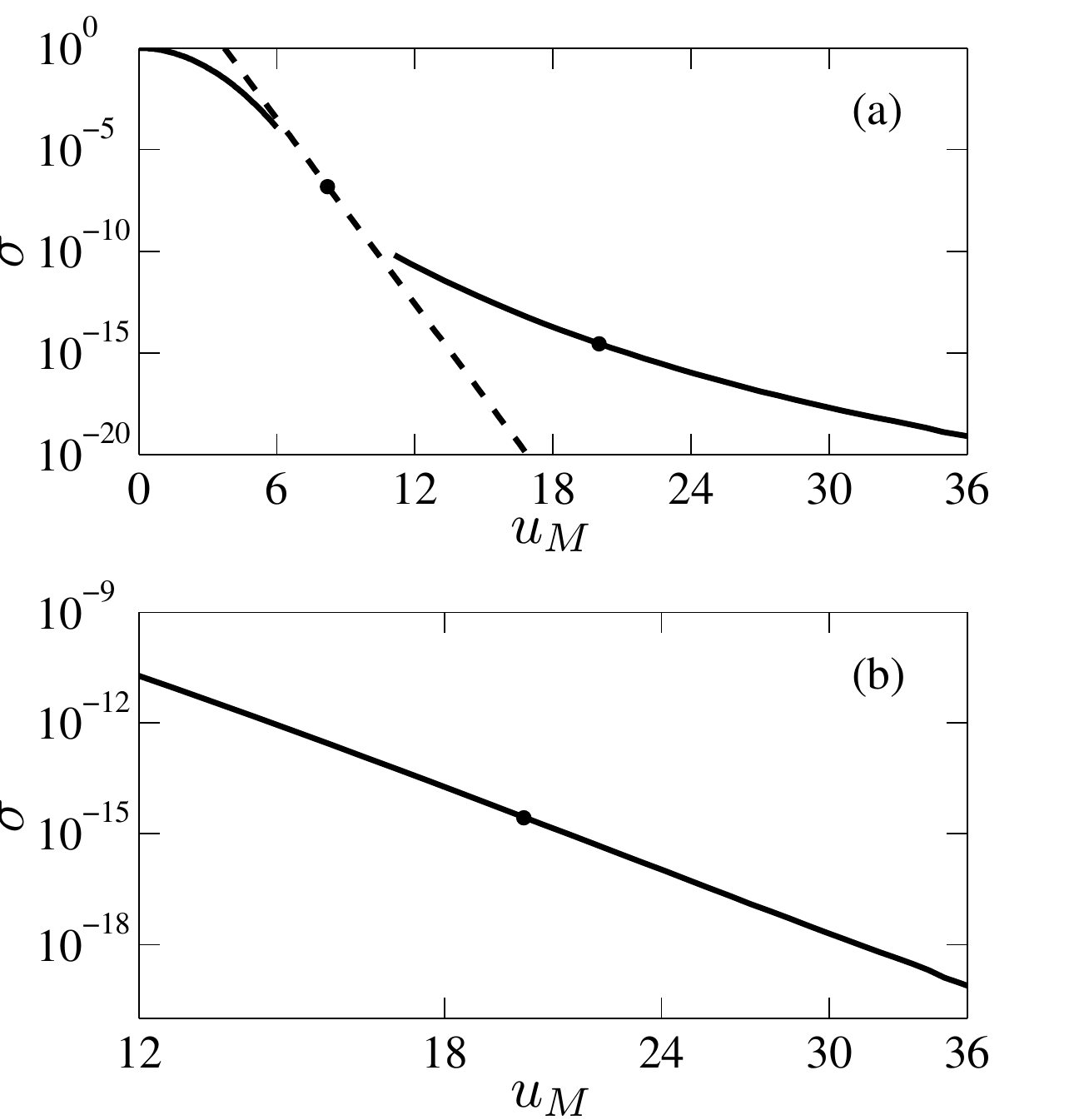}
  \caption{\label{fig:Regimes1D} (a) 1D conductivity \textit{vs.\/} $u_M$ obtained numerically (see main text). The tilted dashed line and the two dots have the same meaning as in Fig.~\ref{fig:1D}. The parabola on the left of the dashed line represents the activation regime. (b) The same data plotted on a double logarithmic scale in order to demonstrate that it can resemble a power-law~\eqref{eqn:G_Power_Law} at low $T$ (to the right of the dot).}
\end{figure}

Using our first fitting formula for $P(u)$ and Eq.~\eqref{eqn:G_avr_I}, we also computed the conductivity at sufficiently large $u_M > 12$ where this approach should be valid. The results are shown in Fig.~\ref{fig:Regimes1D} together with the activated dependence~\cite{Raikh1989fot} at $u_M < 6$ and the dashed line through the surmised inflection point at $u_M = \sqrt{8 c_1}
\approx 8$. The obtained graph suggests a smooth transition from the exponential to the stretched-exponential to the power-law-like behavior of $\sigma(T)$, consistent with Fig.~\ref{fig:1D} and our previous work.~\cite{Rodin2009nso, Rodin2010apl}

\section{Discussion and Conclusion}
\label{sec:Conclusions}

In this paper we have investigated the dependence of VRH conductivity on temperature in systems of finite thickness or length in all physical dimensions. We have demonstrated that both stretched exponential and quasi-power-law dependences can arise in such systems. The significance of our results in elucidating the transition between these different laws.

Prior theoretical work that dealt with hopping conductivity across a thin film took the number of hops $N$ as the principal variable of the problem.~\cite{Tartakovskii1987hco, Glazman1988ita, Raikh1989fot} This allows one to compute the conductivity in the RCH regime beyond the leading logarithmic approximation given in Sec.~\ref{sec:RCH}. However, this approach precludes one from making connection with the percolation theory and the Mott law. Our method, which employs concepts of subcritical percolation theory and takes $u$ as the main variable, allows us to do so. In the RCH regime, the computation of the optimal $u = u_*$ implicitly optimizes the number of hops in a path. In the Mott regime, it plays the role of the percolation parameter directly.

Our results provide a natural explanation for the behavior observed in quasi-1D systems with a large number of channels.~\cite{Zaitsev2000lll,Slot2004odc,Venkataraman2006eti,Zhou2007ode} However, individual wires~\cite{Long2010ecs} and similar systems, like graphene nanoribbons,~\cite{Han2010eti} exhibit significant mesoscopic fluctuations. Further effort is required to include these fluctuations into account.

Examples of 2D systems that can be studied using our method include GaAs devices~\cite{Hughes1996dfa} and bilayer graphene $p$-$n$ junctions.~\cite{Oostinga2007gii} Finally, in 3D our approach describes thin films.~\cite{Xu1995dih, Yoshida1997tah, Colesniuc2010ebo}
Reference~\onlinecite{Yoshida1997tah} indeed reported the crossover from Mott to a power-law-like dependence with decreasing temperature.

In an interesting recent experiment~\cite{Colesniuc2010ebo} the transport across disordered films of Cu-phthalocyanine has been measured over a broad range of temperature $T$ and thickness $L$. A sharp change from strong to weak $T$-dependence below some temperature in the thinnest
films studied therein agrees qualitatively with our theory. The dependence on $L$ is more complicated to analyze because of possible systematic variation of the film morphology and doping level with thickness. 

Extensions of our work may include the study of Coulomb and spin-related effects,~\cite{Bahlouli1994cci, Shahbazyan1994tcr, Ballard2006roc} energy-dependent (e.g., exponential) density of states, and non-Ohmic transport.~\cite{Rodin2010apl, Fogler2005nov}

This work is supported by the University of California Office of President (UCOP) Program on Carbon Nanostructures and ACS of UCSD.

\appendix
\section{Solution of the 1D toy model}
\label{sec:Interp_Form}

Within our toy model, the (unnormalized) probability of connecting the source and drain via $N - 1$ intermediate sites is given by
\begin{equation}
Z_{N-1}(R, L) = g^{N-1}\int_0^R \prod_{i=1}^N dx_{i} \delta(L - \sum_{j=1}^{N}x_j)\,,
\label{eqn:Toy_Model_Prob}
\end{equation}
where $x_i$ is the length of the $i$th hop and $g$ is the density. The normalized probability is obtained by summing over $N$ and multiplying by the factor $e^{-g L}$ known from the Poisson distribution. In order to compute each term we first take its Laplace transform over $L$. This decouples the integrals and yields the
closed-form expression
\begin{equation}
\tilde{Z}_{N-1}(R, s) = g^{N-1}\left(\frac{1-e^{-sR}}{s}\right)^N\,.
\label{eqn:L_transform_Toy_Model_Prob}
\end{equation}
Next, we perform the inverse Laplace transform and sum over $N$. The result is
\begin{equation}
Z(L) = \sum_{N=1}^\infty \sum_{k=0}^N\frac{N(-1)^k[g(L-kR)]^{N-1}}{k!(N-k)!}\Theta(L-kR)\,.
\label{eqn:ZL1}
\end{equation}
In terms of the notations $l = L/R$ and $\rho = gR$ introduced in Sec.~\ref{sec:1D_systems}, the crossing probability becomes
\begin{equation}
P = \sum_{N=1}^\infty \sum_{k=0}^N\binom{N}{k}\frac{(-1)^k[\rho(l-k)]^{N-1}e^{-\rho l}}{(N-1)!}\Theta(l-k)\,,
\label{eqn:Zl1}
\end{equation}
which can be shown to lead to Eq.~\eqref{eqn:P_sol}
by straightforward algebra.

\begin{figure}[h]
  \includegraphics[width=1.5in]{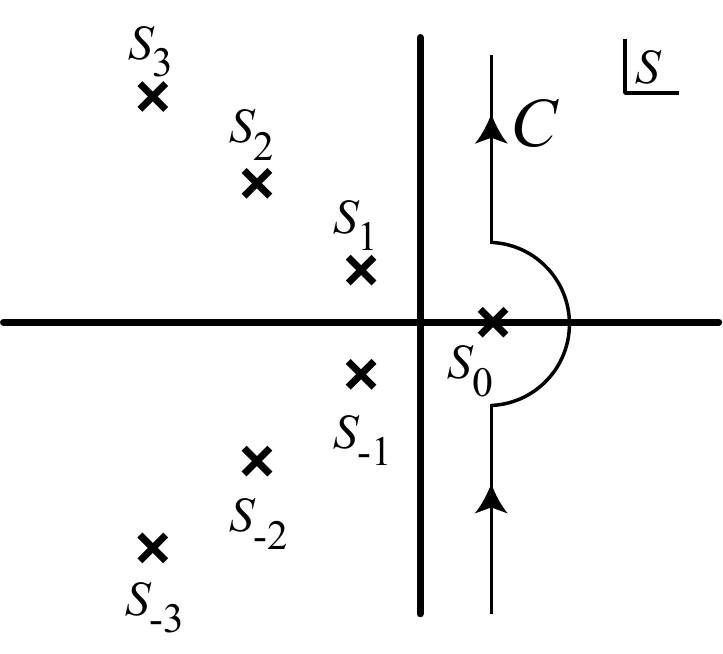}
  \caption{\label{fig:Contour_Integration} The contour used in Eq.~\eqref{eq:Inv_Laplace_Z2}. For $\rho \ll 1$, all the poles $s_j$ are in the left half-plane. When $\rho$ becomes sufficiently large, the pole $s_0$, which has the largest real part becomes positive and approaches $\rho$.}
\end{figure}

One can also proceed by summing Eq.~\eqref{eqn:L_transform_Toy_Model_Prob} for all $N$:
\begin{equation}
\tilde{Z}(R,s) =  \frac{1 - e^{-sR}}{s - g (1 - e^{-sR})}\,.
\label{eq:Total_Prob_Laplace}
\end{equation}
The inverse Laplace transform can be written, after the change of variable $s \to s / R$, as
\begin{equation}
P = \frac{e^{-\rho l}}{2\pi i} \int_C d s\, e^{sl}\, \frac{1 - e^{-s}}{s - \rho (1 - e^{-s})}\,.
\label{eq:Inv_Laplace_Z2}
\end{equation}
The integration contour $C$ is shown in Fig.~\ref{fig:Contour_Integration}.

%

Applying the residue theorem, we get
\begin{equation}
P = \sum\limits_{j=-\infty}^{\infty} \frac{s_j}{\rho (1 - \rho + s_j)}\, e^{(s_j - \rho) l}\,,
\end{equation}
where $s_j$ are the roots of Eq.~\eqref{eqn:k}. For large $l$ the real root $s_0$ dominates; therefore,
%
\begin{equation}
P \simeq \frac{s_0}{\rho(1 - \rho + s_0)}\, e^{(s_0 - \rho) l}\,.
\label{eq:Prob_Misha}
\end{equation}
Equation~\eqref{eqn:Approx_Misha} quoted in Sec.~\ref{sec:1D_systems} was obtained from here by dropping the pre-exponential factor.

\section{Lower bound on the crossing probability}
\label{sec:Low_u}

For any $d$ we define the function
\begin{align}
Z_{N-1} &= \int d \Gamma \delta\left(L - \sum_{k=1}^N x_k \right) \prod_{i=1}^N
 \Theta(u - u_{i, i - 1})\,,
\label{eqn:Z_N-1}\\
d\/ \Gamma &= \prod_{i=1}^N d x_i d^{d - 1} r_i^\perp
            \prod_{j=1}^{N-1} g d\varepsilon_j
\end{align}
Unlike $Z_{N-1}$ in the toy model [Eq.~\eqref{eqn:Toy_Model_Prob}] here we take into account random site energies $-T u < \varepsilon_j < T u$. The energies of the electrode sites, $\varepsilon_0$ and $\varepsilon_N$, are both equal to zero to minimize the resistance. The transverse displacements $\mathbf{r}_i^\perp$ that exist in $d > 1$ case are unrestricted except the first and the last one, $\mathbf{r}_1^\perp = \mathbf{r}_N^\perp = 0$. The hop lengths $x_i$ in the $x$-directions are assumed to be positive.

The sum of $Z_{N-1}$ over all $N$ yields the unnormalized probability of finding a path with $u_{i, i - 1} < u$ for all links $i$ provided only the hops between sites that are nearest-neighbors in $x$ are allowed. Hence, this quantity gives the strict lower bound on the crossing probability $P(u)$. The bound becomes sharp at $u \ll u_M$ where the average number of nearest neighbors is parametrically small.

Similar to Appendix~\ref{sec:Interp_Form}, we proceed to take
the Laplace transform of $Z_{N-1}$ with respect to $L$:
\begin{align}
\tilde{Z}_{N-1} &= \int d\/ \Gamma \prod_{i = 1}^N
\Theta\left(\phi_i - \frac{2}{a} \sqrt{(r_i^\perp)^2 + x_i^2}\,\right)
e^{-s x_i}\,, 
\label{eqn:Z_tilde}\\
\phi_i &= u - \frac{|\varepsilon_i|+|\varepsilon_{i-1}| + |\varepsilon_i-\varepsilon_{i-1}|}{2T}\,,
\label{eqn:phi}
\end{align}
Henceforth we focus on the 1D case. Making two changes of variables, $\zeta \equiv \varepsilon / (u T)$ and $c = a s / 2$, and integrating over $x_i$,
we obtain
\begin{equation}
\tilde{Z}_{N-1} = \left(\frac{a}{2 c}\right)^{N} (g u T)^{N-1}\, f_{N-1}(0)\,,
\label{eqn:Z_tilde_2}
\end{equation}
where $f_0(\zeta) = k(\zeta)$, function $k(\zeta)$ is defined by Eq.~\eqref{eqn:k}, and functions $f_j(\zeta)$ with $j > 0$ obey the recursive integral equations
\begin{align}
f_j(\zeta) &= \int_{-1+\zeta\Theta(\zeta)}^{1+\zeta\Theta(-\zeta)} d\eta\, A(\zeta,\eta)f_{j-1}(\eta)\,,
\label{eqn:f_j}
\\
A(\zeta, \eta) &= 1-e^{-\lambda}\,,
\label{eqn:Kernel1D}
\\
\lambda(\zeta,\eta) &= \left(1-\frac{|\zeta|+|\eta|+|\zeta-\eta|}{2}\right)u c\,.
\label{eqn:lambda}
\end{align}
Similar equations appeared previously in Ref.~\onlinecite{Tartakovskii1987hco}, a near-literal copy of which is available online as Ref.~\onlinecite{Park2000thc}. Following this work,
for $N \gg 1$ we expect $f_{N-1}(\zeta) \simeq \alpha \varkappa^{N-1} \psi(\zeta)$, where $\varkappa =\varkappa(u c)$ is the largest eigenvalue of the integral operator in Eq.~\eqref{eqn:f_j} and $\alpha$ is the overlap of $f_0(\zeta)$ with the corresponding eigenfunction $\psi(\zeta)$. 
This approximation enables us to derive the leading asymptotic behavior of the inverse Laplace transform of the geometric series:
\begin{equation}
Z = \frac{1}{2\pi i}\int_{-i\infty + \gamma}^{-i\infty + \gamma} dc\, \frac{2}{a} \sum\limits_{N = 1}^\infty
\tilde{Z}_{N-1} e^{2 L c / a}\,.
\label{eqn:Z_inverse_Laplace}
\end{equation}
Representing this integral by a sum over the poles of the integrand, we
see that the dominant pole $c_0$ is the real-valued solution of the equation
\begin{equation}
\frac{\varkappa(u c_0)}{u c_0} = \frac{u_M^2}{u^2}\,,
\label{eqn:c_0_sol}
\end{equation}
while the crossing probability is given by Eq.~\eqref{eqn:P_approx}. For $u \ll u_M$ the product $u c_0$ proves to be large and negative. In this limit we can reduce our integral equation to that studied in Ref.~\onlinecite{Tartakovskii1987hco} and obtain the eigenvalue equation in the form
\begin{equation}
\varkappa\simeq -\frac{e^{|u c_0|}}{|u c_0|}\varkappa_T\,,
\quad \varkappa_T \approx 1.18\,.
\label{eqn:kappa}
\end{equation}
Solving this equation in the leading logarithmic approximation yields Eq.~\eqref{eqn:c_0_result}. In order to produce the second analytic fit (the shorter solid line) in Fig.~\ref{fig:Numerics} we solved both the integral eigenvalue problem and Eq.~\eqref{eqn:c_0_sol} numerically. Hence, we are quite confident that what is shown in that Figure is the strict lower bound on $P(u)$.


%

\end{document}